\documentclass[pra,twocolumn,showkeys,superscriptaddress,10pt]{revtex4-1}  
\usepackage{graphicx,epsfig,epstopdf}
\usepackage[dvipsnames]{xcolor}
\usepackage{amsmath}
\usepackage{subfigure}
\usepackage{mathrsfs}
\usepackage{bm}
\usepackage{times}

\begin{document}

\title{High-fidelity geometric quantum gates with short paths on superconducting circuits}
\author{Sai Li}
\thanks{These authors contributed equally to this work.}
\author{Jing Xue}
\thanks{These authors contributed equally to this work.}

\author{Tao Chen}
\affiliation{Guangdong Provincial Key Laboratory of Quantum Engineering and Quantum Materials,
and School of Physics\\ and Telecommunication Engineering, South China Normal University, Guangzhou 510006, China}
\author{Zheng-Yuan Xue}\email{zyxue83@163.com}
\affiliation{Guangdong Provincial Key Laboratory of Quantum Engineering and Quantum Materials,
and School of Physics\\ and Telecommunication Engineering, South China Normal University, Guangzhou 510006, China}

\affiliation{Frontier Research Institute for Physics, South China Normal University, Guangzhou 510006, China}

\date{\today}

\begin{abstract}
Geometric phases are  robust against certain types of local noises, and thus provide a promising way towards high-fidelity quantum gates. However, comparing with the dynamical ones, previous implementations of nonadiabatic geometric quantum gates usually require longer evolution time, due to the needed longer evolution path. Here, we propose a scheme to realize nonadiabatic geometric quantum gates with short paths based on simple pulse control techniques, instead of deliberated pulse control in previous investigations, which can thus further suppress the influence from the environment induced noises.  Specifically, we illustrate the idea  on a superconducting quantum circuit, which is one of the most promising platforms for realizing practical quantum computer. As the current scheme shortens the geometric evolution path, we can obtain ultra-high gate fidelity, especially for the two-qubit gate case, as verified by our numerical simulation. Therefore, our protocol suggests a promising way towards high-fidelity and roust quantum computation on a solid-state quantum system.
\end{abstract}


\maketitle

\section{Introduction}
Quantum computation is based on a universal set of quantum gates \cite{Uni}, including arbitrary
single-qubit gates and a nontrivial two-qubit gate. Recently, superconducting circuits system has shown its unique merits of high designability and  scalability \cite{SC5}, which represents one of the promising  platforms for realizing quantum computer. First, a superconducting transmon device \cite{Tra} is easily addressed as a two-level system, i.e., the ground and first-excited states $\{|0\rangle, |1\rangle\}$  serving as qubit states, which can be operated by a driving microwave field. Second, due to the circuit nature of superconducting qubits, they can be directly  coupled by capacitance or inductance \cite{CP3,CP4,CP5}, for nontrivial two-qubit gate operations.

On the other hand, in quantum computation, the realized quantum gates are  preferred to be robust against the intrinsic errors as well as manipulation imperfections of the quantum systems. To achieve this, realization of quantum gates using the well-known geometric phases \cite{GP1,GP2,GP3} is one of the promising strategies. This is because geometric phases are only relied on the global property of the evolution path in a quantum system and not sensitive to the evolution details, thus provides a promising way towards   robust quantum gates against some local control errors \cite{AN1,AN2,AN3,AN4,AN5,AN6}.

Due to the superiority of geometric phases, geometric quantum computation (GQC) \cite{AA1} and holonomic quantum computation (HQC) \cite{ANA1,ANA2} have been proposed based on the adiabatic evolution. In adiabatic GQC and HQC require that the evolution process in inducing the geometric phases to be  slowly enough so that the nonadiabatic transition among evolution states are greatly suppressed. Therefore, this adiabatic condition results in long evolution time for desired geometric manipulations, where qubit states will be considerably influenced by environment noises, and thus difficult to realize  high-fidelity geometric gates. To remove this main obstacle, nonadiabatic GQC  \cite{NA1,NA2} and HQC \cite{NNA1,NNA2} were then proposed with fast evolutions. These kinds of geometric gates share both merits of geometric robustness and rapid evolution, and thus has attracted much attention. Theoretically, nonadiabatic GQC \cite{NA3,NA4,NA5,NA6} and HQC \cite{NNA3,NNA4,NNA5,NNA6,NNAa1,NNA7,NNA8,NNA9,NNAa2,NNAa3,NNA10,NNA11,NNA12,NNA13,NNAa4,NNA14} have been extensive explored. Meanwhile, nonadiabatic GQC \cite{exp1,exp2,exp3,exp4,AN6} and HQC \cite{ex1,ex2,ex3,ex4,ex5,ex6,ex7,ex8,ex9,ex10,ex11,ex12} have also also been experimentally demonstrated in different quantum systems.

Comparing with nonadiabatic HQC realized with multi-level quantum systems, nonadiabatic GQC (NGQC) can only involve two-level systems, and thus is more easily realized and controlled. However, in previous NGQC schemes, e.g., orange-slice-shaped loops \cite{NA3,NA4,NA5}, to null dynamical phases  from the total phases, they usually require unnecessary long gate-time due to the fact that the evolution states need to satisfy the cyclic evolution and parallel transport conditions. Recently, new conventional/unconventional  NGQC schemes with short evolution paths have been proposed \cite{SNA,toc1,toc2,toc3}, with deliberated and correlated time-dependent parameter control of the govern Hamiltonian, i.e., these schemes need complex pulse control \cite{yuyang}.

Here, to shorten unnecessary long evolution time of previous NGQC without complex pulse control, we propose an approach  to realize universal nonadiabatic geometric quantum gates with short path based on simple pulse control, which is experimentally preferred, {where  short evolution path corresponds to short evolution time under the same driving pulse condition.}  Comparing with previous NGQC schemes, our scheme extends the experimental feasible geometric quantum gates to the case with shorter gate-time, and thus leads to high-fidelity quantum gates as it can reduce the  influence from environment-induced decoherence.  

Moreover, we illustrate our approach by the realization of arbitrary single-qubit and non-trivial two-qubit geometric gates on superconducting quantum circuits system, which is one of the promising platforms for realizing practical quantum computer. Finally, as our scheme is based on simple setup and exchange-only interactions, it is also suitable for many other quantum systems that are potential candidates for physical implementation of quantum computation.  Therefore, our scheme suggests a promising way towards high-fidelity and roust quantum computation on solid-state quantum systems.

\begin{figure}[tbp]
	\begin{center}
		\includegraphics[width=0.8\linewidth]{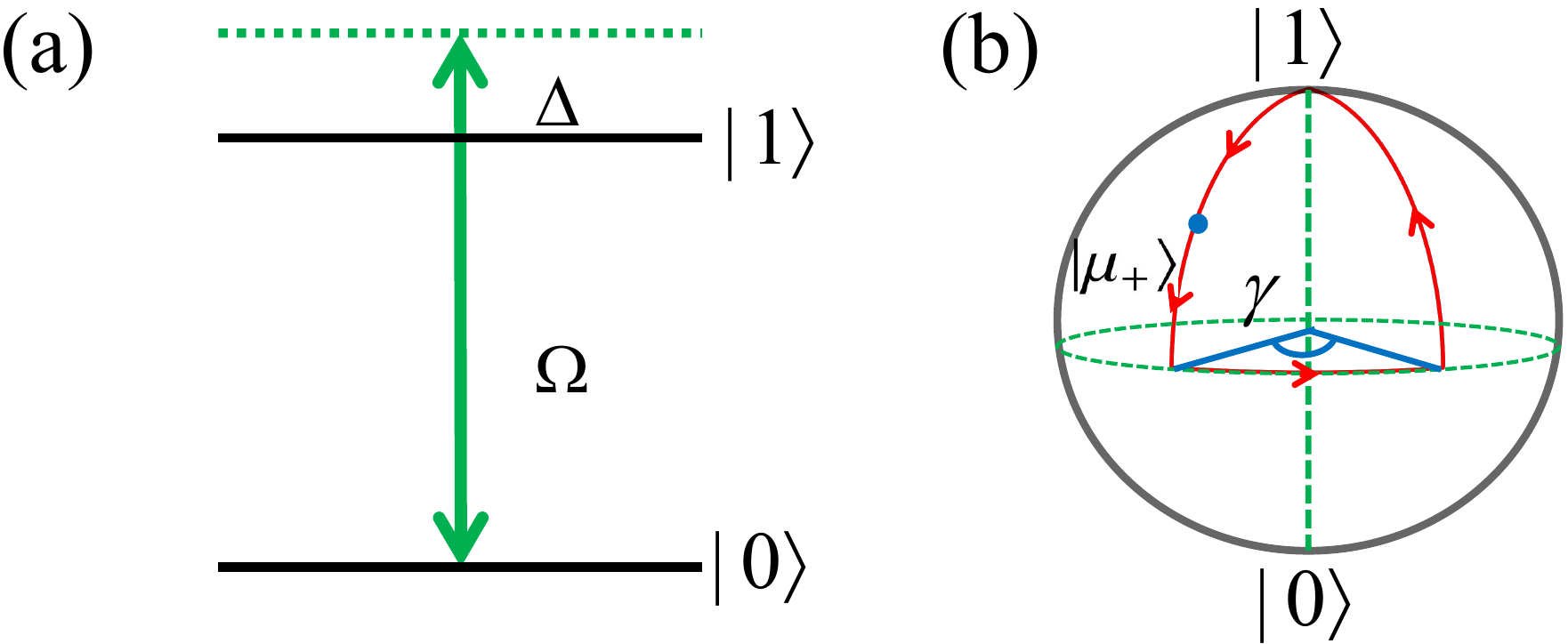}
\caption{Illustration of the implementation of single-qubit geometric quantum gates. (a) The two lowest energy levels of a superconducting transmon qubit can be geometrically manipulated by a  driven microwave field. (b) Geometric illustration of the half-orange-slice-shaped evolution path on a Bloch sphere to induce the geometric phase $-\gamma/2$ on $|\mu_+\rangle$.}\label{F1}
	\end{center}
\end{figure}

\section{Short path geometric gates}

In this section, we  present our general approach to realize nonadiabatic geometric quantum gates with short path based on simple pulse control. We consider a general  setup with a qubit  is driven by a  classical microwave field, as shown in Fig. \ref{F1}(a). Assuming $ \hbar=1 $ hereafter, in the interaction picture and within the computational subspace $\{|0\rangle, |1\rangle\}$, the interaction Hamiltonian of this system is
\begin{eqnarray}
H_{1}(t)=\frac{1}{2}\left(
\begin{array}{cc}
\Delta  & \Omega e^{-i\eta } \\
\Omega e^{i\eta } & -\Delta
\end{array}
\right)\label{H1}
\end{eqnarray}
where $ \Omega $  and $ \eta $  are  the amplitude and phase of the driving microwave field, $\Delta$ is  a frequency difference between the microwave field and the qubit. Here, different with previous investigations, we choose an evolution path as shown in Fig. \ref{F1}(b) to induce our target geometric phases. During this cyclic evolution, which is divided into four parts to acquire a pure geometric phase at final time $T$, parameters of the Hamiltonian is chosen as follows,
\begin{eqnarray}\label{pulse}
&&\int_{0}^{T_{1}}\Omega d t=\frac{\pi}{2}-\theta, \Delta =0, \eta=\phi+\frac{\pi}{2}, t \in[0, T_{1}], \notag\\
&& \Omega = 0,\int_{{T}_{1}}^{T_{2}}\Delta d t= \gamma, \eta=\phi+\gamma-\frac{\pi}{2}, t \in[T_{1}, T_{2}], \notag\\
 && \int_{{T}_{2}}^{T_{3}} \Omega d t=\frac{\pi}{2}, \Delta =0, \eta=\phi+\gamma-\frac{\pi}{2}, t \in[T_{2}, T_{3}], \notag\\
&& \int_{{T}_{3}}^{T} \Omega d t=\theta, \Delta =0, \eta=\phi+\frac{\pi}{2}, t \in[T_{3},T ],
\end{eqnarray}
where parameters $\theta$, $\phi$ and $\gamma$ in each part are easily controlled by external microwave field, {and usually the detuning $\Delta $ is chosen as a fixed constant for easily experimental control purpose}. Then,
at the end of the evolution, the evolution operator can be obtained as
\begin{eqnarray}\label{gates}
U_{1}(T)&&=U_{1}\left(T, T_{3}\right) U_{1}\left(T_{3}, T_{2}\right) U_{1}\left(T_{2}, T_{1}\right) U_{1}\left(T_{1}, 0\right) \notag \\
&&=\cos \frac{\gamma}{2}-i \sin \frac{\gamma}{2}\left(\begin{array}{cc}
\cos \theta & \sin \theta e^{-i \phi}\notag \\
\sin \theta e^{i \phi} & -\cos \theta
\end{array}\right) \\
&&=e^{-i \frac{\gamma}{2} \mathbf{n} \cdot \mathbf{\sigma}},
\end{eqnarray}
which represents rotation operations around axis $\mathbf{n}=\left( \sin \theta \cos \phi ,\sin \theta \sin\phi ,\cos \theta \right) $ with an angle $\gamma$, where $\mathbf{\sigma }=(\sigma _{x}, \sigma _{y}, \sigma _{z})$ are the Pauli matrixes.  Due to phase $\gamma/2$ exactly corresponding to half of the solid angle formed by the evolution path, the evolution operator along this path is naturally induced in a geometric way.  Obviously, this evolution path in our approach is half-orange-slice-shaped loops, which  is shorter than previous nonadiabatic GQC with orange-slice-shaped loops. Meanwhile, during the whole evolution process, the Hamiltonian parameters are only required to meet the conditions as prescribed in Eq. (\ref{pulse}), which can be in arbitrary shape and thus can be chosen as experimental-friendly as possible. That is, our scheme is based on simple and easy pulse control over the  external microwave field, instead of deliberately complex time modulation in previous schemes.

To clearly show the geometric nature of the evolution operator, we give further demonstration of the evolution operator $U_{1}(T)$ as a geometric gate. Here, the two-dimensional orthogonal eigenstates
\begin{eqnarray}
|\mu_{+}\rangle&&=\cos \frac{\theta}{2}|0\rangle+\sin \frac{\theta}{2} e^{i\phi}|1\rangle, \notag\\
|\mu_{-}\rangle&&=\sin \frac{\theta}{2} {e}^{-i \phi}| 0\rangle-\cos \frac{\theta}{2}| 1\rangle
\end{eqnarray}
of $\mathbf{n} \cdot \mathbf{\sigma}$ are selected as our evolution states in dressed representation, after the cyclical evolution, the evolution operator can be expressed in dressed basis $\{|\mu_{+}\rangle, |\mu_{-}\rangle\}$ as
\begin{eqnarray}
U_{1}(T)=e^{-i\frac{\gamma }{2}}|\mu _{+}\rangle \langle \mu _{+}|+e^{i%
\frac{\gamma }{2}}|\mu _{-}\rangle \langle \mu _{-}|.
\end{eqnarray}
It is clear that the orthogonal eigenstates $|\mu_{+}\rangle$ and $|\mu_{-}\rangle$ strictly meet cyclic evolution condition as
\begin{eqnarray}
|\mu _{m}(T)\rangle= U_{1}(T)|\mu _{m}\rangle  = |\mu _{m}\rangle
\end{eqnarray}
with $(m = +,-)$. Meanwhile, the parallel transport condition is also satisfied during the cyclic evolution process as
\begin{eqnarray}
\langle\mu _{m }|U_{1}^{\dag }(t)H_{1}(t)U_{1}(t)|\mu _{m }\rangle
=0,
\end{eqnarray}
which also means that there is no dynamical phases accumulated on the orthogonal eigenstates $|\mu_{m}\rangle$. Therefore, after going through a half-orange-slice-shaped loop, as shown in Fig. \ref{F1}(b), at the end of the evolution, only pure geometric phases are obtained on the orthogonal eigenstates $|\mu_{m}\rangle$ without any dynamic phases. In other words, the total phase is equal to the geometric phase. Thus, arbitrary geometric manipulation can be achieved.

\section{Implementation on superconducting circuits}
In this section, we present the implementation of our short-path NGQC (SNGQC) on superconducting quantum circuits system. We first illustrate the single-qubit implementation with numerically demonstration of the gate performance. By faithful numerical simulations, we also compare our approach against environment-induce decoherence with previous NGQC under the same maximum driving amplitude. Moreover, we compare our approach against different errors with previous NGQC under the same maximum driving amplitude. Then, we proceed to the realization of nontrivial nonadiabatic  geometric  two-qubit gates based on two capacitively coupled transmon qubits, with numerical verifications.

\subsection{Single-qubit gates}
\begin{figure}[tbp]
	\begin{center}
		\includegraphics[width=0.9\linewidth]{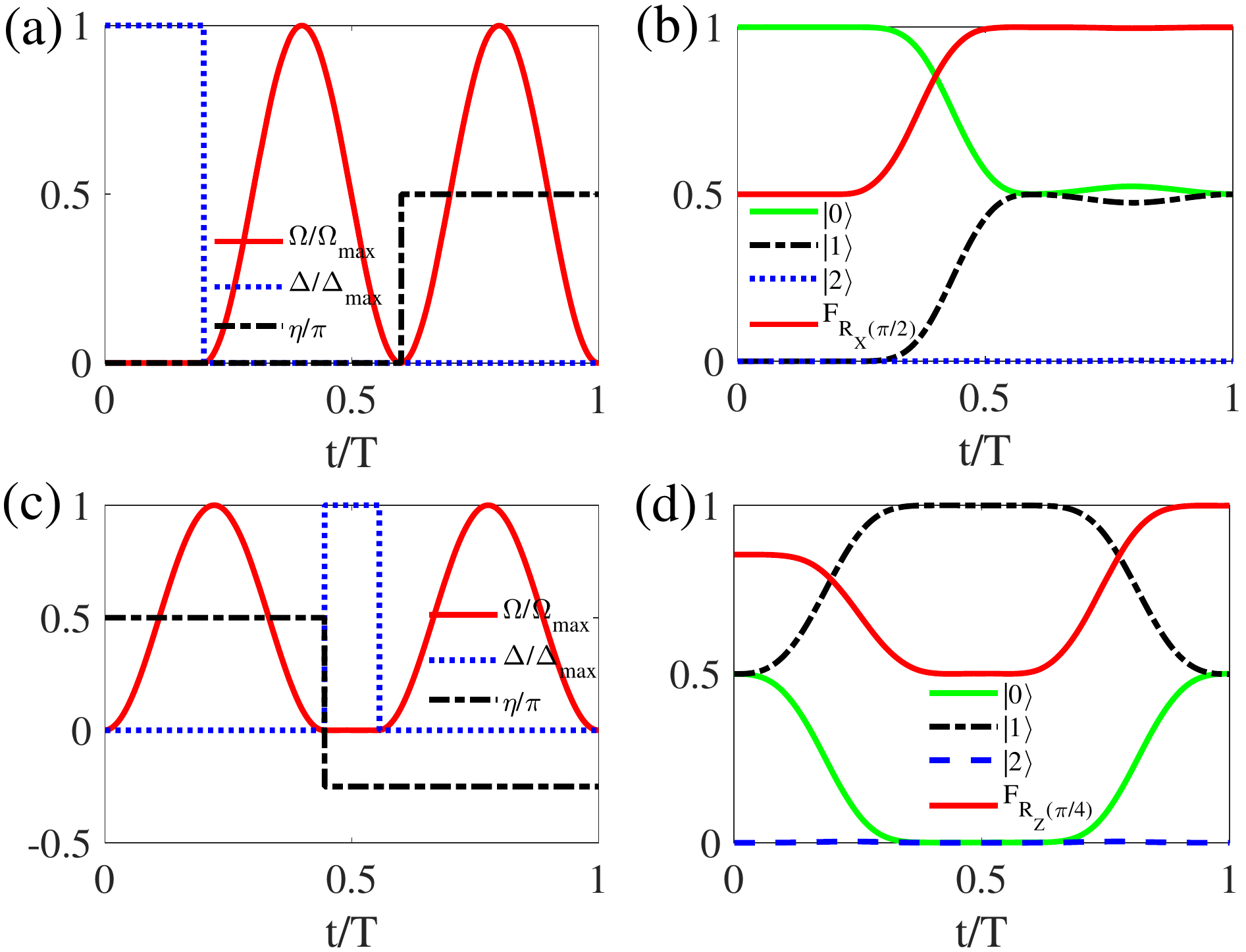}
\caption{Performance of nonadiabatic single-qubit geometric gates. (a) and (c) The shape of parameters $\Omega$, $\Delta$ and $\phi$  for the $R_x(\pi/2)$ and $R_z(\pi/2)$ gates, respectively. (b) and (d) The qubit-state populations and the state-fidelity dynamics of the $R_x(\pi/2)$ and $R_z(\pi/2)$ gate operations, respectively.}\label{F2}
	\end{center}
\end{figure}

Now, we  proceed to the implementation of the nonadiabatic  geometric single-qubit gates on superconducting circuits. In a  superconducting transmon device \cite{Tra},  the ground and first-excited states $\{|0\rangle, |1\rangle\}$  serve as a qubit. A detuned classical microwave field driving, as in Fig. \ref{F1}(a), leads the interacting Hamiltonian as in Eq. (\ref{H1}). Choosing the Hamiltonian parameter according to Eq. (\ref{pulse}), one can obtain the geometric quantum gate as in Eq. (\ref{gates}), which is an arbitrary rotation along the direction set by the parameters $\theta$, $\phi$ and $\gamma$. Thus, arbitrary geometric single-qubit gates are achieved.

\begin{figure}[tbp]
	\begin{center}
		\includegraphics[width=0.9\linewidth]{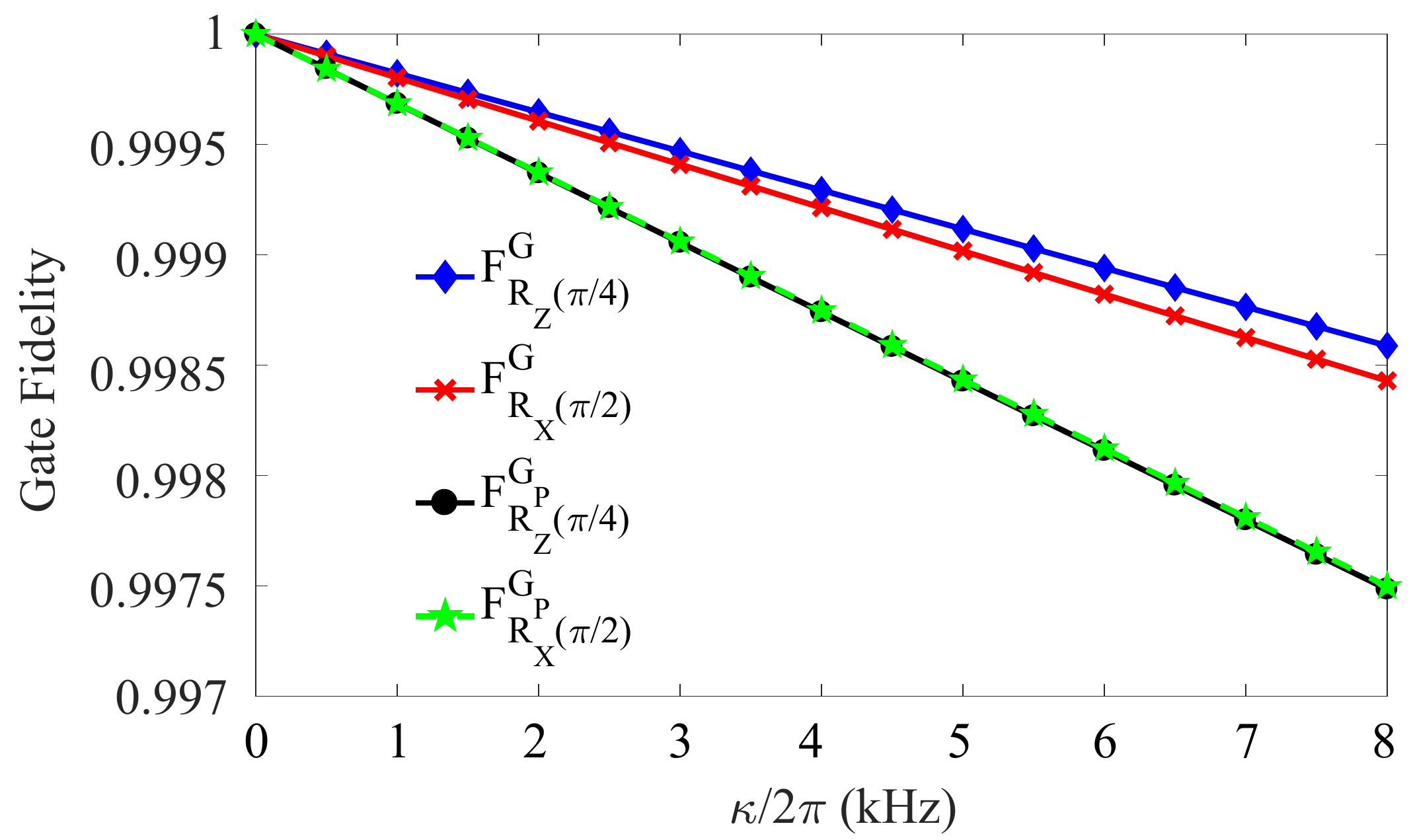}
\caption{Gate fidelities with respect to the different uniform decoherence rate $\kappa$ for the $R_{x}({\pi }/{2})$ and $R_{z}({\pi }/{4})$ gates of our approach and the $R^P_{x}({\pi }/{2})$ and $R^P_{z}({\pi }/{4})$ gates of previous NGQC under the same maximum amplitude.}\label{F3}
	\end{center}
\end{figure}

Then, we numerically demonstrate the performance of our gate performance with faithful numerical simulation.  At first, the influence of environment-induced decoherence in the practical superconducting circuits systems is a non-negligible factor to evaluate gate performance, which is key error in quantum systems, thus, decoherence must be considered for faithful simulation. Meanwhile, as a superconducting transmon device only has  weak anharmonicity, when microwave field is added to drive the transmon qubit with states $\{|0\rangle, |1\rangle\}$, it also drives states $\{|1\rangle, |2\rangle\}$ with $|2\rangle$ being second-excited state, this can cause leakage error from the qubit basis to second-excited state $|2\rangle$. Therefore, the DRAG correction \cite{DR1,DR2} must be introduced to suppress the leakage error beyond the qubit basis. Overall, considering both the decoherence and the leakage term, we introduce the Lindblad master equation as
\begin{eqnarray}
\dot{\rho}_{1}=i \left[\rho_{1}, H_{1}(t)+H_{\textrm{L}}(t)\right]+ \left[\kappa_{1} \mathcal{L}\left(\Lambda_{1}\right)+\kappa_{2}\mathcal{L}\left(\Lambda_2\right)\right],
\end{eqnarray}
with leakage term as
\begin{eqnarray}
H_{\textrm{L}}(t)=(-\alpha-\frac{\Delta}{2}) |2\rangle\langle 2|+[\frac{\Omega}{\sqrt{2}}e^{i\phi}|1\rangle \langle 2|+\mathrm{H.c.}],
\end{eqnarray}
 {where all the unwanted imperfections  are considered by using Hamiltonian $H_{1}(t)+H_{\textrm{L}}(t) $ to faithfully evaluate the gate performance}, $\rho_{1}$ represents the density matrix for the considered system and $\mathcal{L}(\mathcal{O})= \mathcal{O} \rho_{1} \mathcal{O}^{\dagger} -\mathcal{O}^{\dagger} \mathcal{O} \rho_{1}/2-\rho_{1} \mathcal{O}^{\dagger} \mathcal{O}/2$ is the Lindblad operator of $\mathcal{O}$ with $\Lambda_{1}=|0\rangle \langle 1| + \sqrt{2}|1\rangle\langle2|$ and $\Lambda_2=| 1\rangle \langle 1| + 2| 2\rangle \langle 2|$, and $\kappa_{1}$ and $\kappa_{2}$ are the decay and dephasing rates of the transmon qubit, respectively.

\begin{figure}[tbp]
	\begin{center}
		\includegraphics[width=0.9\linewidth]{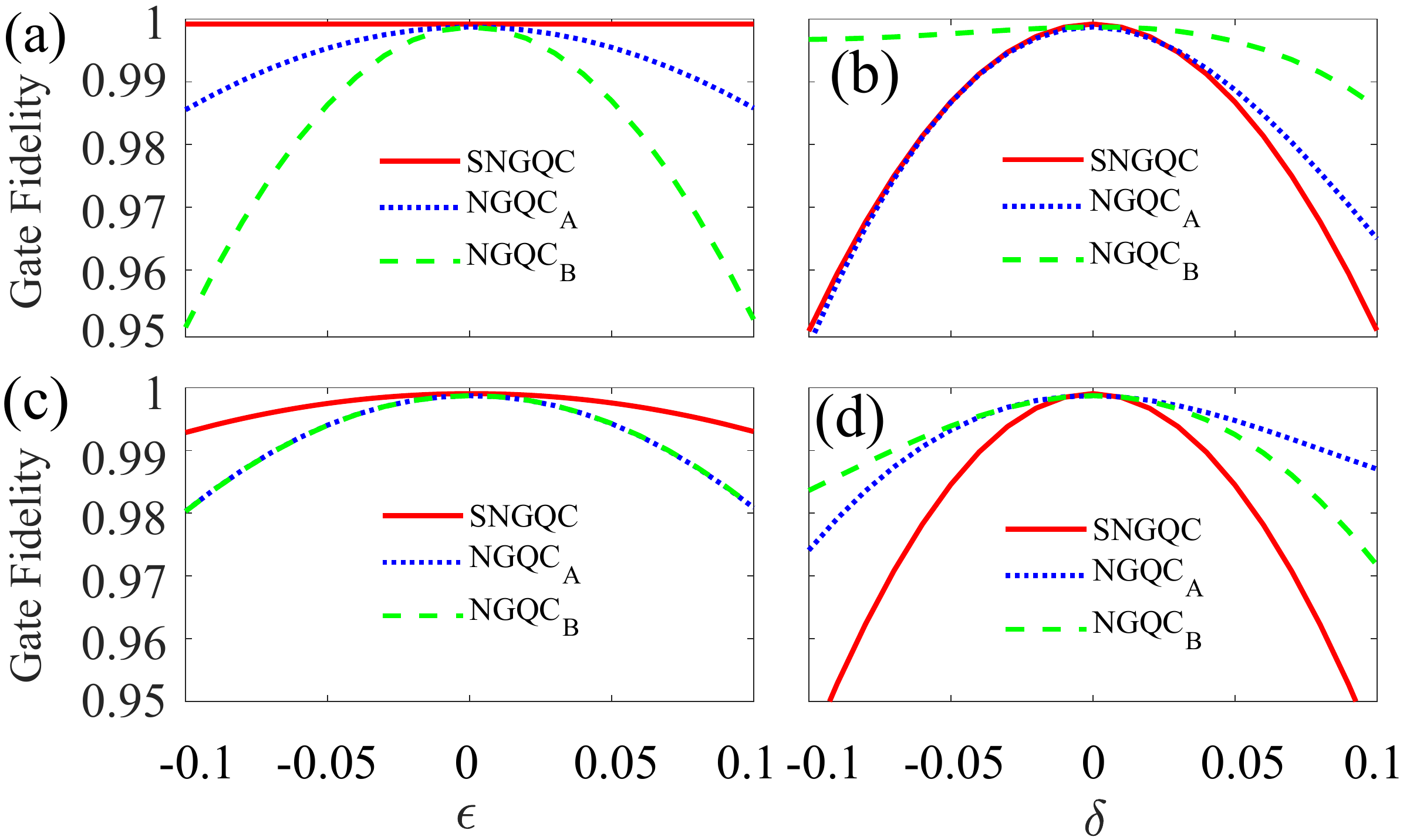}
\caption{Gate robustness comparison. (a) and (c) respectively comparing  the gate robustness for the \emph{S} gate and  \emph{H} gate under the Rabi error $\epsilon$,  (b) and (d) respectively comparing  the gate robustness for the \emph{S} gate and  \emph{H} gate  under the detuning error $\delta$, obtaining from our  SNGQC, NGQC in paths A ($\text{NGQC}_\text{A}$) and  B ($\text{NGQC}_\text{B}$) with decoherence rates $\kappa_{1} = \kappa_{2}= 2\pi\times 4 $ kHz under the same maximum amplitude of the driving field.}\label{F4}
	\end{center}
\end{figure}

Here, to achieve faithful simulation, we set all parameters of the transmon  qubit being easily accessible with current experimental technologies \cite{SC5}, including decay and dephasing rates as $\kappa_{1} = \kappa_{2}= 2\pi\times 4 $ kHz, the weak anharmonicity of the transmon as $\alpha = 2\pi \times 220$ MHz, the maximum amplitude as $\Omega_{\textrm{max}} = 2\pi\times 20$ MHz, and the frequency difference between microwave field and qubit as $\Delta = 2\pi\times 20$ MHz. Meanwhile, due to the weak anharmonicity $ \alpha $ of the transmon qubit, we need to use DRAG correction to suppress the leakage error in order to realize ultra-high gate fidelity, thus, we set the simple form of the driving amplitude as  $\Omega (t)=\Omega _{\textrm{max}}\sin ^{2}(\pi t/\tau)$, where $\tau$ is the duration in each part.

Then, we choose two geometric single-qubit quantum gates $R_{x}({\pi }/{2})$ and $R_{z}({\pi }/{4})$ as typical examples, with gate parameters $\gamma ={\pi }/{2}$, $\theta =\pi/2$, and $\phi =0$ for $R_{x}({\pi }/{2})$ gate and $\gamma = \pi /4$, $\theta =0$, and $\phi =0$ for $R_{z}({\pi }/{4})$ gate. The cyclic evolution time $T$ is about 63 ns for the $R_{x}({\pi }/{2})$ gate and 56 ns for the $R_{z}({\pi }/{4})$ gate. The shape of parameters $\Omega$, $\Delta$ and $\phi(t)$ for the $R_{x}({\pi }/{2})$ and $R_{z}({\pi }/{4})$ gates are shown in Figs. \ref{F2}(a) and \ref{F2}(c), respectively. Assuming the initial states of quantum system are $|\psi(0)\rangle=|0\rangle$ and $|\psi(0)=(|0\rangle+|1\rangle)/\sqrt{2}$ for the $R_{x}({\pi }/{2})$ and ${R_{z}({\pi }/{4})}$ gates, respectively. These geometric gates can be evaluated by using the state fidelity defined by $F=\langle\psi(T)|\rho_1|\psi(T)\rangle$ with $|\psi(T)\rangle=(|0\rangle-i|1\rangle)/\sqrt{2}$ and $|\psi(T)\rangle=(|0\rangle+e^{i\pi/4}|1\rangle)/\sqrt{2}$ being the corresponding ideal final states of the $R_{x}({\pi }/{2})$ and ${R_{z}({\pi }/{4})}$ gates, respectively. The state fidelities are as high as $F _{{R_{x}({\pi }/{2})}}= 99.95\%$ and $F _{{R_{z}({\pi }/{4})}}= 99.88\%$, as shown in Figs. \ref{F2}(b) and \ref{F2}(d), respectively. In addition, for the general initial state $|\psi_1(0)\rangle=\cos\vartheta|0\rangle+\sin\vartheta|1\rangle$, the $R_{x}({\pi }/{2})$ and ${R_{z}({\pi }/{4})}$ gates should result in the ideal final states $|\psi(T)\rangle=(\cos\vartheta-i\sin\vartheta)/\sqrt{2}|0\rangle+(\sin\vartheta
-i\cos\vartheta)/\sqrt{2}|1\rangle$ and $|\psi(T)\rangle=\cos\vartheta|0\rangle+e^{i\pi/4}\sin\vartheta|1\rangle$, respectively. To fully evaluate the gate performance, we define gate fidelity as $F^G_1=(\frac{1}{2\pi}){\int^{2\pi}_0}\langle\psi(T)|\rho_1|\psi(T)\rangle d\vartheta$ with the integration numerically performed for 1001 input states with $\vartheta$ being uniformly distributed over $[0,2\pi]$. We find that the gate fidelities of the $R_{x}({\pi }/{2})$ and ${R_{z}({\pi }/{4})}$ gates can reach as high as $F^G _{{R_{x}({\pi }/{2})}}= 99.92\%$ and $F^G _{R_{z}({\pi }/{4})}= 99.93\%$.

Furthermore, to clearly show the merits of our approach in decreasing the influence of environment-induced decoherence compared with previous nonadiabatic GQC, we depict the trend of gate fidelities under uniform decoherence rate $\kappa/2\pi \in [0,8]$ kHz for the $R_{x}({\pi }/{2})$ and $R_{z}({\pi }/{4})$ gates of our approach and the $R^P_{x}({\pi }/{2})$ and $R^P_{z}({\pi }/{4})$ gates of previous NGQC under the same maximum amplitude $\Omega_{\textrm{max}} = 2\pi\times 20$ MHz, as shown in Fig. \ref{F3}, which directly demonstrate the advantage of our approach over previous ones.

Moreover,  we compare the gate robustness  of our SNGQC scheme for the \emph{S} gate with  gate parameters $\gamma ={\pi }/{2}$, $\theta =0$, and $\phi =0$ and  \emph{H} gate with  gate parameters $\gamma ={\pi }$, $\theta =\pi/4$, and $\phi =0$ with previous NGQC in both paths A and B (see Appendix A for details) at different error conditions, including the Rabi error case of $H^\epsilon_{1}(t)=\frac{1}{2}[\Delta\sigma_z+ (1+\epsilon)(\Omega e^{-i\eta }|0\rangle\langle1|+\text{H.c.})]$ with different Rabi errors  $\epsilon \in[-0.1,0.1]$ and the detuning error case of $H^\delta_{1}(t) = H_{1}(t) +\delta\Omega_\text{max}|1\rangle\langle1|$ with different  detuning errors $\delta \in[-0.1,0.1]$. Here, we simulate above process with the decoherence rates $\kappa_{1} = \kappa_{2}= 2\pi\times 4 $ kHz under the same maximum amplitude $\Omega_{\textrm{max}} = 2\pi\times 20$ MHz, and the results are shown in Fig. \ref{F4}. Due to the different evolution paths between SNGQC and NGQC respectively dominated by different Hamiltonian, the Hamiltonian in NGQC case is not commuted with detuning error, however, the Hamiltonian in SNGQC has $\sigma_z$ component,  these can lead to the different asymmetry and symmetry fidelity behavior respect to zero detuning point with regard to NGQC and SNGQC case in Figs. \ref{F4}(b) and (d). While our SNGQC scheme does not performs well under the detuning error $\delta$ than previous NGQC in both path A and B, as shown in Figs. \ref{F4}(b) and (d),  our SNGQC scheme does perform better under the Rabi error $\epsilon$, as shown in Figs. \ref{F4}(a) and (c). To sum up, our  SNGQC scheme can share the merits of decreasing the influence of environment-induced decoherence and the gate robustness against the Rabi errors than previous NGQC, and thus is promising for quantum systems where the $X$ error is dominated.

\begin{figure}[tbp]
	\begin{center}
		\includegraphics[width=0.9\linewidth]{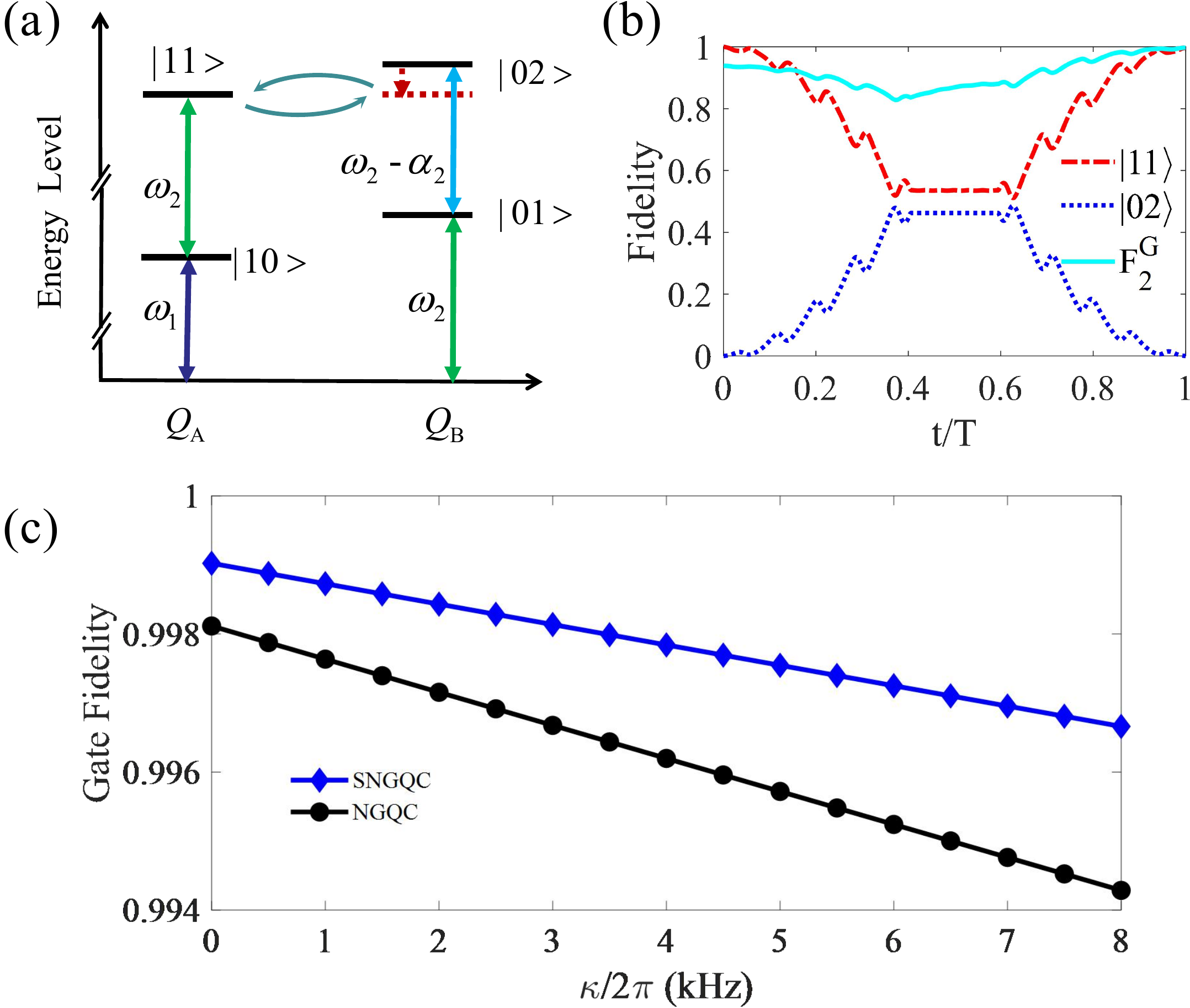}
\caption{Illustration of the realization of the geometric two-qubit
 gates. (a) Energy level of coupled transmon qubits $Q_A$ and $Q_B$. (b) State dynamics and the gate fidelity of a nontrivial geometric control-phase gate with $\gamma^\prime =\pi/2$. (c) Gate fidelities with respect to the different uniform decoherence rate $\kappa$ for the $U_2 (\pi/2)$ gate of our approach and previous NGQC under the same maximum Rabi amplitude. }\label{F5}
	\end{center}
\end{figure}

\subsection{Nontrivial two-qubit gates}

We further proceed to introduce the realization of nontrivial nonadiabatic  geometric  two-qubit gates with short path based on two capacitively coupled transmon qubits \cite{CP3,CP4,CP5}, labeled as $Q_A$ and $Q_B$ with qubit frequency $\omega_{{A},{B}}$ and anharmonicity $\alpha_{{A},{B}}$. Usually, the frequency difference $\zeta=\omega_\textrm{B}-\omega_{A}$ and coupling strength $g$ between these two transmon qubits $Q_A$ and $Q_B$ are fixed and not adjustable. To achieve tunable coupling and desired interaction between them \cite{CP3,CP4,CP5}, an ac driving can be added on the transmon qubit $Q_B$, which results in periodically modulating the frequency of $Q_B$ as
$\omega_B(t) = \omega_B+\epsilon \sin(\nu t+\varphi)$.
Then, in the interaction picture, the Hamiltonian of coupled system can be expressed as
\begin{eqnarray}\label{HC}
{H}_C(t)&&=g[|01\rangle_{{AB}}\langle10|e^{i\zeta t }+ \sqrt{2}|11\rangle_{{AB}}\langle20|e^{i(\zeta+\alpha_{A}) t } \notag\\
&&+\sqrt{2}|02\rangle_{{AB}}\langle11|e^{i(\zeta-\alpha_{B}) t }]e^{-i\beta\cos(\nu t + \varphi)}+\textrm{H.c.},
\end{eqnarray}
where $\beta = \epsilon/\nu$ and $|jk\rangle=|j\rangle\otimes|k\rangle$. Here, as shown in Fig. \ref{F5}(a), only the interaction in the subspace $\{|11\rangle_{{AB}}, |02\rangle_{{AB}}\}$ is considered  by choosing the driving frequency $\nu=\zeta-\alpha_{B}+\Delta^\prime$ with $g\ll\{\nu, \zeta-\nu, \zeta+ \alpha_{A}-\nu \}$, and then using Jacobi-Anger identity $\exp[i\beta\cos(\nu t + \varphi)]=\sum^\infty_{n=-\infty}i^nJ_n(\beta)$\\
$\exp[in(\nu t + \varphi)]$ with $J_n(\beta)$ being the Bessel function of the first kind, and neglecting the high-order oscillating terms, the obtained effective Hamiltonian can be reduced to
\begin{equation}\label{Heff}
\begin{aligned}
H_{2} = \frac {1} {2} \left(\begin{array} {cc} {\Delta^\prime} & {g^\prime e^{i \eta^\prime}} \\ {g^\prime e^{-i \eta^\prime}} & {-\Delta^\prime} \end{array} \right),
\end{aligned}
\end{equation}
in the two-qubit subspace $\{|11\rangle_{{AB}}, |02\rangle_{{AB}}\}$, where $g^\prime = 2\sqrt{2}gJ_1(\beta)$ is effective coupling strength between transmon qubits $Q_A$ and $Q_B$, $\Delta^\prime$ is the energy difference between states $|11\rangle_{{AB}}$ and $|02\rangle_{{AB}}$, and $\eta^\prime=\varphi+\pi/2$.

Then, the Hamiltonian $H_{2}$ can be directly applied to acquire a pure geometric phase $e^{-i\gamma^\prime/2}$ on two-qubit state of $|11\rangle_{{AB}}$ by a cyclic evolution beyond the computation basis like the way of constructing geometric single-qubit rotation operations $R_z(\gamma)$ around axis $\sigma_z$. Thus, within the two-qubit computation subspace $\{|00\rangle_{{AB}}, |01\rangle_{{AB}}, |10\rangle_{{AB}}, |11\rangle_{{AB}}\}$, the final nontrivial geometric two-qubit control-phase gates can be acquired as
\begin{equation}\label{U2}
\begin{aligned}
U_2 ( \gamma^\prime ) =  \left(
\begin{array}{cccc}
1 & 0 & 0 & 0 \\
0 & 1 & 0 & 0 \\
0 & 0 & 1 & 0 \\
0 & 0 & 0 & e^{-i\frac{\gamma^\prime}{2}} \\
\end{array}
\right).
\end{aligned}
\end{equation}

{Here,  we use Hamiltonian ${H}_C(t)$ in Eq. (\ref{HC})  considering all the unwanted imperfections to faithfully evaluate the nontrivial two-qubit geometric control-phase gates,} and we also apply the Lindblad master equation  with $\gamma^\prime = \pi/2$ as a typical example. Then, to achieve faithful simulation, we also set all parameters of the transmon qubits being easily accessible with current experimental technologies \cite{SC5}, including the frequency difference $\zeta =2\pi\times 500$ MHz, anharmonicity of qubits $\alpha_{A} =2\pi\times 220$ MHz and $\alpha_{B} =2\pi\times 230$ MHz, $g = 2\pi\times 10$ MHz, the driving frequency $\nu = \zeta-\alpha_{B}+\Delta^\prime=2\pi\times 270$ MHz with $\Delta^\prime=0$ and  $\nu = \zeta-\alpha_{B}+\Delta^\prime=2\pi\times 300$ MHz with $\Delta^\prime=2\pi\times 30$ MHz, effective coupling strength of $g^\prime_{max}\approx2\pi\times 14$ MHz with $\beta =1.2$, and the decoherence rate of transmons being the same as the single-qubit case \cite{SC5}. Then, the cyclic evolution time T for the two-qubit gate is about $44$ ns. The state dynamics of subspace $\{|11\rangle_{{AB}}, |02\rangle_{{AB}}\}$ are verified in Fig. \ref{F5}(b). To faithfully evaluate the gate performance of two-qubit gates, for the general initial state  $|\psi_2(0)\rangle = (\cos\vartheta_1 |0\rangle_{A} + \sin\vartheta_1 |1\rangle_{A} )\otimes(\cos\vartheta_2 |0\rangle_{B} + \sin\vartheta_2 |1\rangle_{B} )$ with $|\psi_{2}(T)\rangle  = U_2(\pi/2)|\psi_2(0)\rangle$ being the ideal final state, we can define the two-qubit gate fidelity as
\begin{equation}
\label{TG}
F^G_2 = \frac{1}{4\pi^2}\int^{2\pi}_0\int^{2\pi}_0\langle\psi_{2}(T)|\rho_2|\psi_{2}(T)\rangle d\vartheta_1 d\vartheta_2,
\end{equation}
with the integration numerically done for 10001 input states with $\vartheta_1$ and $\vartheta_2$ uniformly distributed over $[0, 2\pi]$. As shown in Fig. \ref{F5}(b), we can obtain the gate fidelity $F^G_2=99.78\%$, where the infidelity is caused by decoherence about $0.1\%$ and leakage errors about  $0.1\%$.

{In addition, we  also simulate the trend of two-qubit gate fidelities under uniform decoherence rate $\kappa/2\pi \in [0,8]$ kHz for the $U_2 (\pi/2)$ gate of both our and previous NGQC approaches, under the same maximum amplitude $g^\prime_{max}=2\pi\times 14$ MHz, as shown in Fig. \ref{F5}(c), which also demonstrate the advantage of our approach over previous ones. Notably, under the fixed parameters anharmonicity of qubits $\alpha_{A} =2\pi\times 220$ MHz, $\alpha_{B} =2\pi\times 230$ MHz, $g = 2\pi\times 10$ MHz, and the same maximum amplitude, we can only  optimize the frequency difference $\zeta =2\pi\times 490$ MHz to suppress the leakage errors to $0.2\%$ for  previous NGQC.}

\section{Conclusion}
In summary, we propose an approach to realize universal nonadiabatic geometric
quantum gates with short path based on simple pulse control to shorten unnecessary long evolution time of previous MGQC without complex pulse control. {In addition, our scheme can perform better under the Rabi errors than NGQC with orange-slice loops.} Our approach extends experimental feasible geometric quantum gates with shorter evolution process to futher reduce the influence of environment-induced decoherence compared with previous NGQC. Meanwhile, our approach is suitable for many quantum physical systems, e.g., superconducting circuits systems. We further demonstrate our approach by the realization of arbitrary single-qubit geometric gates and non-trivial two-qubit geometric gates on superconducting circuits systems.

\appendix
\section*{Appendix}
\section{Previous NGQC with orange-slice loops}
{In this appendix, we present the details in implementing  previous  NGQC  with orange-slice-shaped loops in path A and Path B. In path A, the orange-slice-shaped evolution path is divided into three parts with resonant driving by microwave drive with the amplitude $ \Omega $ and phase $\eta$, which satisfy }
\begin{eqnarray}\label{path A}
&&\int_{0}^{T_{1}}\Omega d t=\theta,  \eta=\phi-\frac{\pi}{2}, t \in[0, T_{1}], \notag\\
 && \int_{{T}_{1}}^{T_{2}} \Omega d t={\pi}, \eta=\phi-\gamma+\frac{\pi}{2}, t \in[T_{1}, T_{2}], \notag\\
&& \int_{{T}_{2}}^{T} \Omega d t={\pi}-\theta, \eta=\phi-\frac{\pi}{2}, t \in[T_{2},T ],
\end{eqnarray}
{then, the geometric evolution operator can be obtained as $U_A(T) = e^{-i\gamma  \mathbf{n} \cdot \mathbf{\sigma}}$. And in path B, the geometric evolution is realized by setting $\eta=\phi-\gamma-\frac{\pi}{2}$ at $\in[T_{1}, T_{2}]$, while the corresponding  geometric evolution operator keeps the same form of $U_A(T)$.  However, these two geometric paths have different gate robustness against different type of errors show in Fig. \ref{F4} in the main text. }

\section*{Acknowledgements }
This work was supported by the Key-Area Research and Development Program of GuangDong Province (Grant No. 2018B030326001), the National Natural Science Foundation of China (Grant No. 11874156), the National Key R\&D Program of China (Grant No. 2016 YFA0301803), and the Science and Technology Program of Guangzhou (Grant No. 2019050001).

\end{document}